\documentclass[useAMS,usenatbib]{mn2e}
\usepackage{graphicx}

\title[CoRoT-7 planet(s)]{Re-assessing the radial-velocity evidence for planets around CoRoT-7}
\author[Pont, Aigrain \& Zucker]{Fr\'ed\'eric Pont$^{1}$\thanks{E-mail:
email@address}, Suzanne Aigrain$^{2}$, Shay Zucker$^{3}$\\
$^{1}$School of Physics, University of Exeter, EX4 4QL, Exeter,  UK\\
$^{2}$Astrophysics, Department of Physics, University of Oxford, OX1 3RH, Oxford, UK\\
$^{3}$Department of Geophysics and Planetary Science, Tel Aviv University, 69978 Tel Aviv, Israel}
\begin{document}

\date{Accepted 2010 October 5.  Received 2010 October 5; in original form 2010 August 24}

\pagerange{\pageref{firstpage}--\pageref{lastpage}} \pubyear{2010}

\maketitle

\label{firstpage}

\begin{abstract}
CoRoT-7 is an $11^{th}$ magnitude K-star whose light curve shows transits with depth of 0.3\,mmag and a period of $0.854$\,d, superimposed on variability at the 1\% level, due to the modulation of evolving active regions with the star's 23\,d rotation period. In this paper, we revisit the published HARPS radial velocity measurements of the object, which were previously used to estimate the companion mass, but have been the subject of ongoing debate. 

We build a realistic model of the star's activity \emph{during the
HARPS observations}, by fitting simultaneously the line width (as
measured by the width of the cross-correlation function) and the line
bisector, and use it to evaluate the contribution of activity to the
RV variations. The data show clear evidence of errors above the level
of the formal uncertainties, which are accounted for neither by
activity, nor by any plausible planet model, and which increase rapidly with
decreasing signal-to-noise of the spectra. We cite evidence of similar
systematics in mid-SNR spectra of other targets obtained with HARPS
and other high-precision RV spectrographs, and discuss possible
sources. Allowing for these, we re-evaluate the semi-amplitude of the
CoRoT-7b signal, finding $K_b=1.6\pm1.3$\,m\,s$^{-1}$, a tentative
detection with a much reduced significance (1.2-sigma) compared to previous
estimates. We also argue that the combined presence of activity and
additional errors preclude a meaningful search for additional low-mass
companions, despite previous claims to the contrary.

Taken at face value, our analysis points to a lower density for CoRoT-7b, the $1\,\sigma$ mass range spanning $1$--$4\,M_{\rm Earth}$, and allowing for a wide range of bulk compositions. In particular, an ice-rich composition is compatible with the RV constraints. More generally, this study highlights the importance of a realistic treatment of both activity and uncertainties, particularly in the medium signal-to-noise ratio regime, which applies to most small planet candidates from CoRoT and Kepler.\end{abstract}


\section{Introduction}

\subsection*{First of a kind}
 
%Presentation of CoRoT-7
 
CoRoT-7 is an active K-dwarf which was monitored photometrically for 5
months in 2007--2008 as part of the exoplanet programme of the CoRoT
space mission \citep{Bag03}. Following the routine analysis of the
data to search for planetary transits, \citet[][hereafter L09]{Leg09}
reported the detection, in CoRoT-7's light curve, of eclipses with a
depth of 0.3\,mmag and a period $P_b=0.854$\,d, superimposed on
significant (1.8\% peak-to-peak) activity-induced stellar
variability. They also carried out ground-based photometric follow-up
and obtained near-infrared spectra, which ruled out the majority of
the alternative binary scenarios that could have given rise to
the observed transits (grazing or diluted eclipsing systems with a
stellar, sub-stellar or giant planet companion). They therefore
interpreted the eclipses as most likely to be caused by a planetary
companion, dubbed CoRoT-7b, with a radius of $R_b =1.8\pm0.2\,
R_{\oplus}$. The accompanying radial velocity (RV) follow-up campaign,
carried out with the HARPS spectrograph, was reported in
\citet[][hereafter Q09]{Que09}. The RV signal of CoRoT-7 was dominated
by strong (40\,m\,s$^{-1}$  peak-to-peak) activity-induced variations. Q09
nonetheless derived a radial velocity semi-amplitude of $K_b
=3.5\pm0.6$\,m\,s$^{-1}$, corresponding to a mass
$M_b=4.8\pm0.8\,M_{\oplus}$. They also reported the detection of
another (non-transiting) planet in the system, CoRoT-7c, with period
$P_c=3.69\,$d, semi-amplitude $K_c=4.0\pm0.5$\,m\,s$^{-1}$, and mass
$M_c=8.4\pm0.9\,M_\oplus$.

%Importance of CoRoT-7. Up to now, only significant inroad towards the stated goal of the mission - habitable $\sim2 R_earth$ planets (Leger REFS). Although modest: very short period, brightest star.

Optimistic pre-launch estimates of CoRoT's detection capabilities had
forecast the detection of hundreds of transiting planets, with masses
between that of Neptune and twice that of the Earth \citep[see
e.g.][]{Bor03}. However, prior to CoRoT-7b, all the planets that
CoRoT discovered were gas giants. The smallest transiting planet
known was the Neptune-mass GJ\,436b \citep{Gil07}, whose transits were
detected from the ground after it was discovered by radial
velocities. Thus, the discovery of CoRoT-7b was an important
confirmation that space-based transit surveys could indeed detect
planets in this regime. It should be noted that CoRoT-7 was a
particularly favourable target: it is one of the brightest targets
ever monitored as part of CoRoT's exoplanet programme ($R=11.4$), and
the precision of its CoRoT light curve is photon-noise limited on the
2-h timescales typical of short-period transits \citep{Aig09}, despite
significant variability on longer time-scales.

%Previous literature on CoRoT-7. 

%Scientific issues, position relative to other Neptune-mass transiting planets.

A handful of new low-mass transiting planets have since been detected: GJ\,1214\,b, another `Super-Earth', by the ground-based MEarth project \citep{Cha09}, and two `hot Neptunes', Kepler-4b \citep{Boru10} and HAT-P-11b \citep{Bak10}. However, such objects remain scarce and their potential to constrain formation and evolution scenarios unique. As a result, the CoRoT-7b system has already been the subject of a number of theoretical studies. The mass and radius of CoRoT-7b, as reported by Q09, point to a predominantly rocky composition \citep{Val10,Bar10}, albeit with significant degeneracy \citep{Rog10}. This is in stark contrast to GJ\,1214b, whose lower density indicates a significant H$_2$O content and/or H/He envelope \citep{Cha09}, and it raises interesting questions about CoRoT-7b's origin \citep{Jac09,Jac10}. \citet{Dvo10} also investigated the dynamics of the two-planet system reported by Q09, pointing out signatures of possible interactions between the two planets which may be observable in the medium term. All of these studies depend critically on the assumed parameters of the planet(s) and host star. 

The scientific importance of this object, and the extremely challenging nature of the observations, have naturally prompted a number of teams to re-analyse the HARPS RV data presented in Q09. Far from clarifying the situation however, these efforts appear to be leading to an even more confused picture. Before summarising them, it is helpful to summarise briefly the methodology of Q09.

\subsection*{Previous analyses}

%Assessment of previous literature. Several studies, but all with basically the same approach, and sub-sets of the same initial CoRoT team. 
 
Q09 first analysed the RV data using a pre-whitening procedure,
successively fitting and subtracting sinusoids at the period
corresponding to the highest peak in a Lomb-Scargle periodogram. They
attributed the first three peaks, at 23.5, 9.03 and 10.6\,d, to
activity, on the basis that they are close to the rotation period of
the star (determined to be $P_{\rm rot} \sim 23.2\,$d from the CoRoT
photometry) or one of its first few harmonics. The fourth peak is not
obviously related to $P_{\rm rot}$, and they attributed it to a planet
which does not transit (CoRoT-7c). Finally, the fifth peak was found
at the one-day alias of the period of CoRoT-7b (though it should also
be noted that this corresponds almost exactly to the third harmonic of
the stellar rotation period, $P_{\rm rot}/4$). In support of the
planetary origin of this signal, Q09 point out that its phase is
consistent with the CoRoT ephemeris. As an alternative to the
pre-whitening procedure, they also modelled the activity signal by
fitting a sum of three sinusoids at $P_{\rm rot}$, $P_{\rm rot}/2$ and
$P_{\rm rot}/3$ to the data. Because the active regions evolve rapidly
(this is manifested both in the light curve and in the RV data
itself, whose scatter changes significantly from observing run to
observing run), this `3-harmonic filter' was adjusted using a sliding
window of duration $18$\,d, which effectively introduces many
more degrees of freedom to the fitting process. In both cases, they
later fit a two-planet Keplerian model to the residuals, fixing
the period and phase of the transiting planet to the CoRoT
ephemeris. Having established that both filtering procedures could
reduce the amplitude of a putative planetary signal at the period of
CoRoT-7b by as much as 50\%, they corrected for this by 
multiplying the amplitude derived from the filtered data by a factor
two. Finally, they averaged the results of the two approaches to
obtain a final semi-amplitude for each planet, but adopted final
uncertainties which are smaller than those derived from either
method. This implicitly assumes that the two approaches give
independent estimates of the semi-amplitudes, which is not the case.

\citet{Hat10} re-analysed the HARPS radial velocity (RV) data using a pre-whitening procedure similar to that of Q09, finding results marginally consistent with Q09 for the masses of CoRoT-7b and c, but interpreting the signal at 9.03\,d as being due to a third planet. This interpretation was not supported by \citet{Lan10b}, who modelled the host star's activity by fitting a spot model to the CoRoT light curve, finding evidence for significant differential rotation. The signal at 9.03\,d is then readily interpreted as the second harmonic of the rotation period. \citet{Lan10b} also simulated the activity-induced RV signal corresponding to their best-fit spot model, which they then resampled at the sampling of the HARPS observations. They used multiple realisations of these simulated time-series to confirm that the `3-harmonic' filter of Q09 adequately removes most of the activity signal (though it would be surprising if it did not, given its large number of free parameters), and to put upper limits of $\sim 0.9$ and $\sim 1.7$\,m\,s$^{-1}$ on the activity-induced signal which survives the 3-harmonic filter at the periods of CoRoT-7b and c respectively. However, the CoRoT and HARPS observations were not simultaneous but separated by one year. Given the rapid evolution of the active regions on CoRoT-7, one cannot be certain that the RV time-series simulated by Lanza et al. is representative of the real activity signal at the time of the HARPS observations. 

\citet{Bru10} also re-determined the fundamental parameters of the host star CoRoT-7 by re-analysing the HARPS spectra of Q09, and re-fit the CoRoT light curve using the updated stellar parameters, resulting in a slightly smaller radius estimate for CoRoT-7b, as well as smaller formal errors ($R_c=1.58\pm0.10\,R_\oplus$), which exacerbates the apparent discrepancy between the compositions of CoRoT-7b and GJ\,1214b. 

\subsection*{Our approach}

%Our approach: also use the simultaneous temperature and bisector information. Real activity as opposed to idealised. Understand the effect of "multi-sine" approach. Consider also the possibility of HARPS instrumental systematics.

%REF for non sum-of-sine-ness and high variability of stellar activity and solar activity. REFS on successful modelling in terms of activity features modulated by rotation. REFS on features of actual activity curves of CoRoT (Suzanne). REFS on the presence of HARPS systematics.

The goal of this paper is to re-assess the mass constraints on the CoRoT-7 planet(s), with some key differences to previous studies with similar goals. 

First, we make use of all available information in our analysis. The
data derived from the HARPS spectra consist not only of radial
velocities, but other parameters derived from the cross-correlation
function (CCF), particularly the bisector span and the full-width at
half-maximum (FWHM), of the CCF. Both are activity diagnostics:
rotating spots affect the wings of the spectral lines and hence modify
the bisector \citep[][hereafter B09]{Boi09}, and the FWHM because it
is a diagnostic of the stellar photospheric temperature \citep{San02},
which is affected by changes in the spotted fraction of the stellar
disk (since the total luminosity and radius are constant).  Q09
discussed those indicators, as well as the Ca {\sc ii} H \& K
chromospheric activity index, but did not directly use them. Yet any
realistic model for the activity-induced RV variations must also
account for the behaviour of these parameters.

Second, we avoid the use of pre-whitening and harmonic filtering
techniques. The sampling of the HARPS data is very irregular,
consisting of some isolated points, some runs of a week or so with one
point per night, and some more intensive runs with up to three points
per night, spanning more than four months in total. The unique
decomposition of arbitrary functions into sines and cosines, which
forms the basis for Fourier analysis and techniques such as
pre-whitening, does not apply to very irregularly sampled data, when
sines and cosines no longer form an orthogonal basis. Filtering out
specific frequencies in the CoRoT-7 HARPS data unavoidably affects
other frequencies, including those corresponding to the planetary
orbital periods of interest, and it is extremely difficult to quantify
the uncertainty in the final semi-amplitudes. This is particularly
acute in the case of CoRoT-7b, whose orbital period (or rather its
alias, since the period itself is above the approximate Nyquist
limit of the observations) is very close to the third harmonic of the
stellar rotation period. Furthermore, the typical lifetimes of active
regions on Sun-like stars are comparable with their rotation periods
(typically $1$--$4\,P_{\rm rot}$, \citealt{Mos09}). As a result, the
frequency content of the activity is highly complex, implying that
pre-whitening or harmonic filters must either be extended to a large
number of frequencies, or be made adaptive, and in any case have a
very large number of free parameters.

On the other hand, we wish to construct a realistic model of the
activity of the star \emph{during the HARPS observations}. Spot models
can reproduce the light curves of the Sun \citep{Lan03,Lan04} and of
CoRoT active planet-host stars \citep{Mos09,Lan09a,Lan10a} to a
high degree of accuracy, as well as the RV curves of active stars
\citep[see e.g.][ B09]{Bon07}. In spite of this success, those
models suffer from strong degeneracies between, for example, the
contrast, size and number of spots and facular areas. More recently,
spot models built from solar magnetograms were used to produce forward
models of the photometric and RV variations of the Sun, which are in
excellent agreement with the observations
\citep{Lag10,Meu10a,Meu10b}. However, in the case of CoRoT-7, we have
already seen that the CoRoT light curve is of limited use because it
is not simultaneous with the RV data. Instead, we build our activity
model from the HARPS data itself, using the CCF FWHM as a photometric
proxy. Q09 demonstrated a very tight correlation between the FWHM and photometry obtained simultaneously with the 1.2m Euler
telescope, though they did not use it directly in their analysis.  We
model this proxy photometry using large numbers of spots, with minimal
assumptions on their properties. Because of the inherent degeneracies
in spot models, it is standard practice to limit the number of spots
to a few, but typical CoRoT light curves are more representative, in a
statistical sense, of many-spot models (see Aigrain et al.\ in
prep.). Following a maximum entropy logic, we use multiple
realisations which fit the data well to account for, rather than
attempt to circumvent, the degeneracy inherent in many-spot models.

Finally, we also consider the possibility of systematics in the RV data, i.e.\ of noise not accounted for by the formal errors. HARPS is optimized and calibrated primarily for ultra-high precision RV monitoring of very bright ($V<8$) stars, and its stability down to the m\,s$^{-1}$  level is well established in that magnitude range \citep[see e.g.][]{May09}. On the other hand, the HARPS uncertainties become significantly larger, and highly dependent on factors such as moonlight and weather, for faint stars, as seen during the follow-up of $V \sim 15$ OGLE transit candidates \citep{Bou05}. As we will discuss in Section~\ref{syst}, this trend is seen at low signal-to-noise ratio across a wide range of target types and instruments. At $V=11.7$, CoRoT-7 falls in the intermediate SNR regime for HARPS, and the possibility of systematics should at least be considered.

\section{The activity signal}

\subsection*{Characteristics of activity-induced signals}

Stellar activity induces photometric variations, primarily via the rotational modulation of dark spots on its surface. Although there are exceptions, these star spots typically have lifetimes on the scale of one to a few stellar rotation periods \citep{Mos09}. As a result, the lightcurves of active stars are often highly complex, although they usually display some periodicity at the rotation period or a low-order multiple thereof.

Star spots also have an effect on RV measurements. The first-order effect is due to the spots occulting either the receding or approaching half of the visible hemisphere, thus modifying the shape of the spectral lines and causing a shift in the apparent disk-averaged RV. The difference in temperature and convective upwelling velocity between the spots and the rest of the surface of the star also has an effect, which will depend on the wavelength range used to measure the RVs, and the spectral lines included in the measurement. Again, the activity-induced RV `jitter', as it has come to be known, can be very complex and hard to interpret, partly because of spot evolution and partly because of the irregular time sampling of ground-based RV data.

%Apart from slow-evolving surface features modulated by rotation, stellar activity also causes features on shorter timescales, such as solar flares.  

Aside from the Sun, one of the stars with the best studied RV jitter is HD\,189733, which was the target of intensive simultaneous photometric and RV monitoring, with the MOST satellite and SOPHIE instrument respectively (B09). Like CoRoT-7, HD\,189733 is an active K-dwarf with a close-in planetary companion, although the star is much brighter and the planet much more massive. B09 showed that simple models of the rotational modulation of dark spots account for most of the observed photometric and RV variations of HD\,189733 (once the planetary signal is removed), as well as for other activity diagnostics such as the CCF bisector span and emission in the cores of the Ca {\sc ii} H \& K and H$\alpha$ lines. The bisector span, which measures the mean RV offset between the core and the wings of spectral lines, appears to be a particularly robust jitter indicator. B09 used the correlation between RV and bisector span to correct the SOPHIE RV time-series of HD\,189733 for the effects of activity down to the $\sim 9$\,m\,s$^{-1}$  level, at which point the RV data became limited by instrumental systematics.

We have undertaken an investigation of stellar activity in the CoRoT
light curves, which will form the subject of a forthcoming paper
(Aigrain et al.\ in prep.). One point from that study which is
particularly relevant to the present work is the fact that typical
CoRoT light curves are better described by models including many spots
($20$--$200$) than a few ($2$--$3$). Few-spot models can reproduce
specific cases, but in general they tend to predict time
intervals of near-constant flux, corresponding to times when no large
spot is present on the visible hemisphere. Such intervals are almost
completely absent from the CoRoT light curves, which suggests that, as
in the case of the Sun\footnote{See monthly average sunspot numbers as
tabulated by the Solar Influences Data Analysis Center, {\tt
http://www.sidc.oma.be}.}, the surfaces of Sun-like stars may be
peppered with tens to hundreds of small spots, rather than a few large
ones.

\subsection*{Activity modelling for CoRoT-7}

%\begin{figure}
%\resizebox{8cm}{!}{\includegraphics{x.png}}
%\caption{}
%\label{}
%\end{figure}

\begin{figure}
\includegraphics[width=\linewidth]{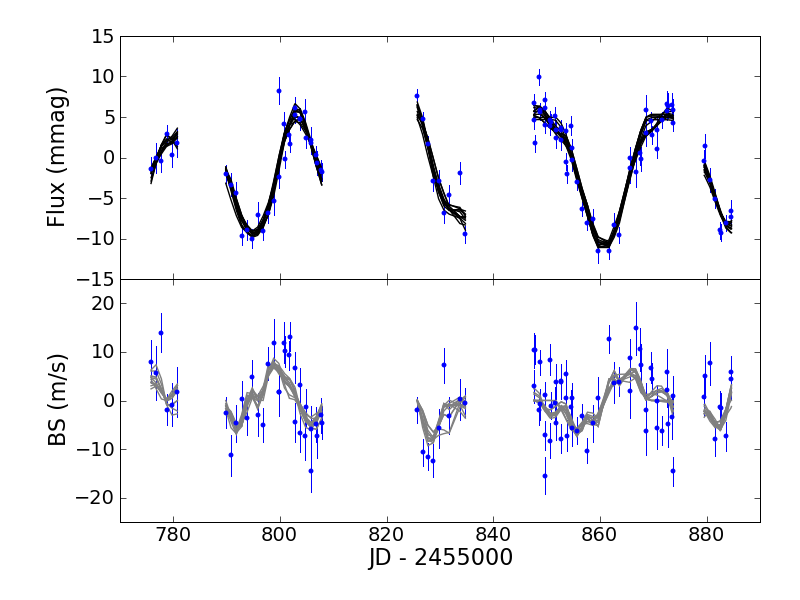}
%\resizebox{12cm}{!}{\includegraphics{fig1.png}}
\caption{{\bf Top:} Proxy light curve for CoRoT-7 inferred from the HARPS CCF width data. The lines show ten models using 12 to 200 rotating star spots, with differential rotation and varying lifetimes (see text for details). {\bf Bottom}: Line bisector data from HARPS spectra, and predictions from the ten spot models. In this example, the bisector data is \emph{not} used to constrain the spot models.}
\label{fig1}
\end{figure}

\subsubsection*{Activity of CoRoT-7}

The CoRoT lightcurve of CoRoT-7 shows variability at the percent
level, with a rotation period of $\sim 23$ days, and rapidly-evolving
activity features (no feature is reproduced unchanged after one
rotation period, and the lightcurve becomes unrecognisable after
merely 2--3 periods). The RV data reported in Q09 show variations at
the 20\,m\,s$^{-1}$  level. The dominant periodicity changes from run to run
between $P_{\rm rot}$, $P_{\rm rot}/2$ and $P_{\rm rot}/3$, which
again implies rapid evolution of the activity features. Similar
variations are observed in the accompanying activity indicators
(bisector span and Ca {\sc ii} H \& K index), although they are more
sensitive to measurement errors.

\begin{figure}
\includegraphics[width=\linewidth]{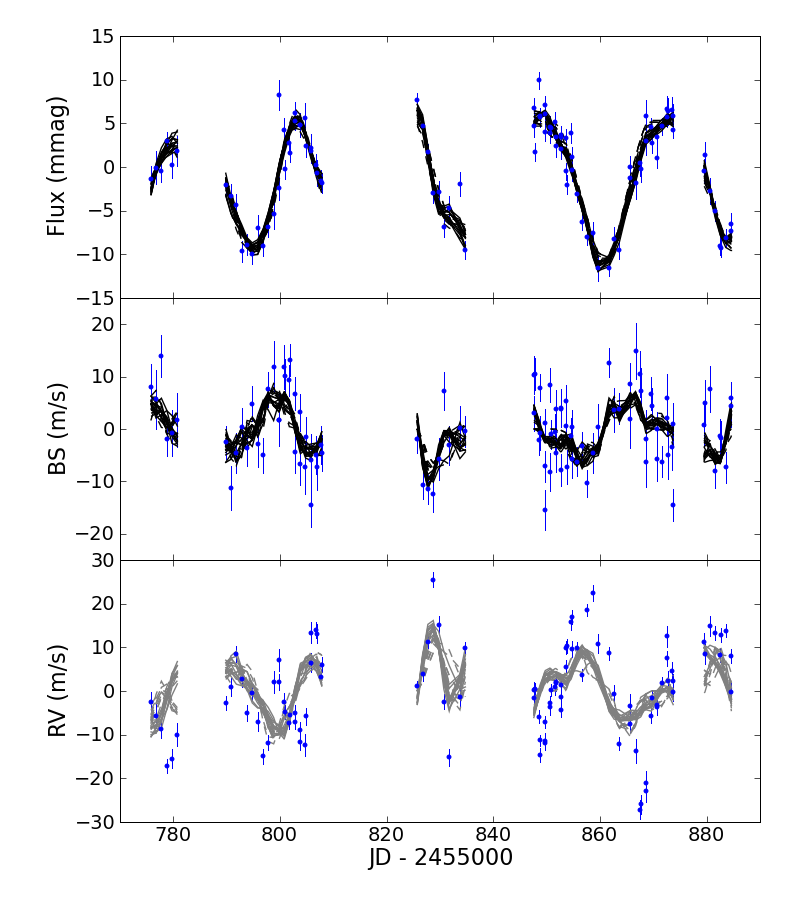}
%\resizebox{12cm}{!}{\includegraphics{fig2.png}}
\caption{From top to bottom: proxy light curve, radial velocities and line bisector span data from the HARPS spectra of CoRoT-7. The lines show fourteen models using 12 to 25 rotating star spots, with differential rotation and varying lifetimes  (see text for details). The models were fitted to both the light curve and line bisector information. %Dotted lines show models where more weight was given to the bisector information. 
The RV variations predicted by the models are shown overlaid on the data in the lower panel.}
\label{fig2}
\end{figure}

We do not have direct observations of the photometric variability of
CoRoT-7 at the time of the HARPS observations. However, Q09
demonstrated the existence of a tight, linear correlation between the
HARPS CCF FWHM and simultaneous photometry obtained with the Euler
1.2-m telescope. This correlation implies that the FWHM can be used to
reconstruct the brightness variations to a very impressive precision,
0.1\% or better. We thus used the FWHM data and conversion relation
from Q09 to reconstruct the light curve of CoRoT-7 during the HARPS
observations. The results are shown in the top panels of Figures~\ref{fig1} and
\ref{fig2}, alongside the RV and bisector span data which were
directly taken from Q09. The amplitude and dominant periodicity of the
variations in the reconstructed light curve are similar to those
observed by CoRoT nearly a year earlier, although the variations are
more sustained in the former. The RV data clearly show
significantly more variability on timescales of a few days and below
than either the proxy photometry or the CoRoT light curve. %RV jitteris generally concentrated on higher harmonics of the rotation periodthan intrinsic brightness variations, but the difference is still noticeable.

Q09 did not directly make use of this proxy photometry. However, it is very constraining. Any self-consistent model for the RV variations in terms of activity plus planetary signals must also account for the FWHM and bisector span variation. Adaptive harmonic or pre-whitening filters such as those used by Q09 and \citet{Hat10} can readily account for rapid RV variations, but unless those coincide with corresponding features in the proxy photometry or bisector span data, they cannot safely be assumed to be due to activity.

\subsubsection*{Our rotating time-evolving spot model}

We model the activity signal with a set of $N$ spots, each
characterised by a scale factor $f$, which is equal to the fraction of
the stellar flux it would `hide' if it  were at the centre of the
stellar disk. To allow for spot evolution, the scale factor varies
with time according to a Gaussian function  (to account for growth
and decay of the spot area). The Gaussian function is parameterised by
the peak intensity $f_0$, peak epoch $t_0$, and lifetime $\tau$ (the
half-width at half-maximum of the Gaussian). Each spot has longitude
$\delta$, and rotates with the surface of the star. The effect of the
spots on the stellar brightness, RV and bisector span is simulated
using a simple analytical model described in Aigrain et al. (in
prep.). Although we treat each spot as point-like, the effect of large
active regions can be readily reproduced with several spots
close together. We use a linear limb-darkening law with $u=0.6$
\citep[e.g.][]{Sin10},  and assume that the star's equator coincides
with the line of sight. The effects of faculae and convective
blueshift are not included\footnote{the good agreement of predicted
and observed bisector information suggest that the spot shading effect
is dominant.}.

We use the proxy photometry, and optionally the bisector span data, to
constrain the spot parameters, and predict the activity-induced
variations in bisector span (if not used to constrain the fit)
and RV. The spot parameters $f_0$, $t_0$, $\tau$, $\delta$ and
$\phi_0$, are set at random at the beginning of each realisation, and
a fit to the data is found by gradient minimisation of the $\chi^2$
statistic. The number of spots $N$ can vary between 3 and 200 (but is
kept fixed for any given realisation). $t_0$ and $\delta$ are free to
vary linearly, and $f_0$ and $\tau$ logarithmically. We allow a
certain amount of differential rotation, leaving each spot to have its
own rotation period  $P=(1+\epsilon)P_0$, where $P_0$ is the equatorial
rotation period. This differential rotation is constrained by adding a
penalty term $(P-P_0)^2/\sigma_P^2$ to the $\chi^2$.

 For each value of $N$, there are many possible realisations that are compatible with the data. There is generally a value of $N$ below which the actual lightcurve cannot be reproduced within the uncertainties, and a value above which adding new spots does not make a visible difference. We find that in the case of CoRoT-7, spot numbers between 12 and 20 reproduce the light-curve within the uncertainties. With less than 6 spots, flat spot-free regions are left in the model light-curve that are not visible in the reconstructed one. We run the downhill minimisation for $10^6$ steps, and accept the fit if the residuals are lower than twice the adopted values of $\sigma_f$ and $\sigma_{bs}$. By varying the initial parameters for the spots and running many realisations, we aim to explore the most likely regions in parameter space. This approach contrasts with models using a low number of large spots, which are finding only one possible inversion of the data, and is similar in spirit to the maximum-entropy reconstruction method used for instance in \citet{lanz10}. We verify that the values of $\sigma_f$ and $\sigma_{bs}$ include not only random noise but also the uncertainties in the modelling procedure, so that the model will not tend to use additional spots to over-fit the observed light-curve and bisector data. 

 Based on the spot configuration, the photometric signal is easily
obtained through a simple integration of the flux over the visible
hemisphere. The RV and bisector signals are a little more complex and are obtained
through an integration which takes into account the Doppler shift of
each point of the visible stellar surface \citep[e.g.][]{dal06}\footnote{The correlation between the RV signal and the bisector variation can be used as a first-order correction to the RV data \citep{Boi09}. However, the relation between RV and bisector is not tight:  a large spot close to the centre of the stellar disc has the same RV effect as a small spot close to the edge but a very different bisector signature. As a result the RV-bisector correlation will be complex \citep[e.g.][]{san00,san03} and a more sophisticated model of the variability is desirable when attempting to study a signal of lower amplitude than the activity signal.}.

\subsubsection*{Results}

To test the suitability of the model, we first fit the light curve
data only and use it to predict the bisector span variations.
Fig.~\ref{fig1} shows the results of ten representative realisations,
with $N$ varying from 12 to 200, $P_0=23.64$ days and $\sigma_P$ set
to 5\%. The agreement with the observed bisector span variations is
remarkable. This indicates that the spot model is essentially
adequate, and that the effects which we neglected (faculae and
convective blueshift) are of secondary importance. A few detailed
features vary from one model to the next, but overall all realisations
with similar goodness of fit give essentially identical photometric
and bisector span variations, regardless of $N$. In other words, one
should not read too much into the individual spot parameters of each
model, but the predictions in terms of photometric and spectroscopic
variations are very robust, particularly over the stretches with
relatively tight and regular sampling.

We also estimated the variations in Ca {\sc ii} H \& K expected for a given spot model.
We take the expected strength of the Ca {\sc ii} H \& K signal to be proportional to the visible spot surface. We find this indicator to be compatible with the flux and bisector data, but significantly noisier. As a result, the comparison does not add much in the way of constraining information, and it is not included in the rest of the analysis.

\subsubsection*{Predicting the activity-induced RV variations}

We repeated the fits using both the proxy light curve and the bisector span data to constrain the spot model, and predicted the corresponding RV variations for each realisation. To do this, one must define a relative weight between the photometry and bisector span data. We used inverse variance weighting with $\sigma_{\rm f} = 0.1 $\% for the light curve and $\sigma_{\rm bs}=3$\,m\,s$^{-1}$  for the line bisector as the default values, as well as $P_{\rm rot}=23.64$ days and $\sigma_P$=1.2 days. The values  of $\sigma_{\rm f}$ and $\sigma_{\rm bs}$ were chosen to reflect the level at which we believe the spot model can be regarded as accurate. We explored the effects of varying these assumptions by computing solutions with $P_{\rm rot}$ from 22.0 to 25.0 days, $\sigma_{\rm f} = 0.3 $\%, and without using the bisector information. 

Figure~\ref{fig2} shows the result of 14 different realisations, covering a wide range of $N$ and spot parameters, with the corresponding RV predictions. There are always multiple good fits to the light curve and bisector data, whatever the value of $N$ and the relative weighting given to the two constraining datasets. The RV predictions reproduce the overall features of the RV data quite well, showing that activity is the dominant source of RV signal for CoRoT-7. However, there are notable departures from our predictions: features in the RV data that are clearly not accounted for by any of the activity models. The question that remains to be answered is: what is the origin of these features?

\section{Analysing the non-activity-induced RV signal}

\subsection*{Residual RV features}

\begin{figure*}
\resizebox{12cm}{!}{\includegraphics{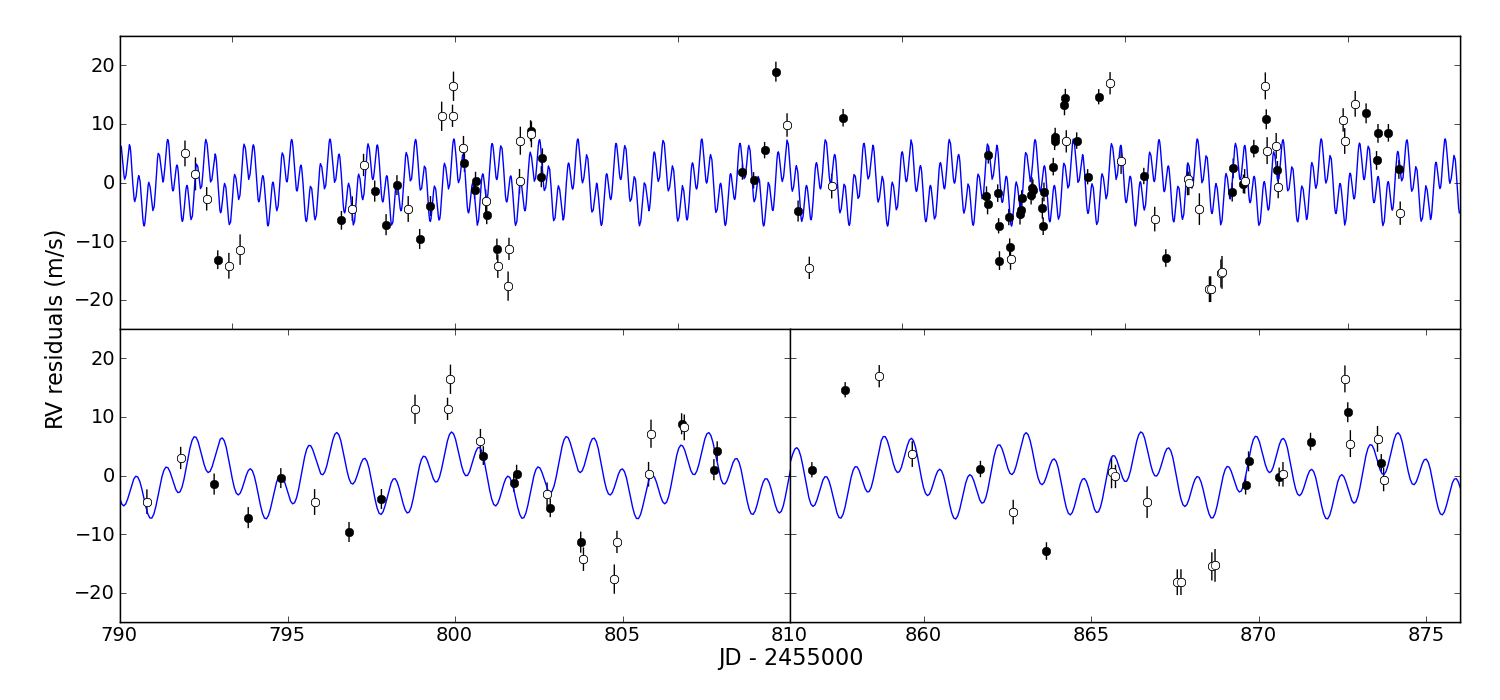}}
\caption{Radial velocity residuals relative to the median of the activity models, compared to the signal expected from the 2-planet model of Q09. Open symbols indicate measurements with lower signal-to-noise ratios. The bottom panels are zoomed in on two specific time intervals. Note in particular the $\sim$\,20 m\,s$^{-1}$  jumps near ${\rm JD}-2455000=795$ and 806, and the set of low points at ${\rm JD}-2455000=868$ and 869. These are not predicted by the activity model or by the two-planet solution, and are much larger than the formal RV uncertainties. }
\label{fig4}\label{fig3}
\end{figure*}

%The "steps". Not rotation of activity feature. Not planets. 

Figure~\ref{fig4} shows the RV data with the median of our activity
model curves subtracted. Two different colours
have been used to distinguish measurements based on spectra with low
and high signal-to-noise ratio (SNR). Data with SNR lower than the
median are displayed in white. The bottom panels of Fig.~\ref{fig3} zoom
in on the two most densely-covered periods, and overplots the expected
planetary signal from the two-planet solution in Q09.

%Probably instr. syst. Could also be stellar feature, but sudden and reversible (unlikely). Anyway the effect on planet detection would be the same.  

Figure~\ref{fig3} shows that the RV data contain features that cannot
be accounted for by the activity signal responsible for the light
curve and bisector variations, nor by the two-planet model of Q09. Of
particular note are the $\sim 20$\,m\,s$^{-1}$  jumps at ${\rm
JD}-2455000=788$, 806 and 868. These jumps occur over the space of one
day, with no corresponding feature in the light curve and bisector
data. It is very difficult to imagine how activity could cause such a
sudden change in RV without leaving a detectable signature in the
photometric or bisector span data\footnote{It is of course conceivable
that such an effect exists, but in that case it is indistinguishable
from instrumental systematics, and effectively plays the same role, so
we will treat it as such.}.  Since the semi-amplitude of the two
putative planets is $4$\,m\,s$^{-1}$ at most, they could not produce
jumps of this amplitude either, even if they were perfectly in phase
with them.

%If such steps and "recovery" are present, lower-level, intermediate-period stability doubtful. Claim of sub-m\,s$^{-1}$  precision on one of many components, including 3 non-periodic component, given 20 m\,s$^{-1}$  non-understood systematics, is fantasy.
 
Figure~\ref{fig3} also shows that the features in the RV residuals are
strongly correlated with the SNR of the spectra. This suggests that
the variations unaccounted for by activity are caused by uncertainties
beyond the formal error bars, increasing with lower SNR. As discussed
in the introduction and in the next section, there are reasons to
believe that such uncertainties may be present in HARPS data, at the
required level of $\sim 5 $\,m\,s$^{-1}$ r.m.s. The SNR is principally
determined by the observing conditions. Observing conditions at
observatories vary strongly over one night for a given object, as the
airmass changes with the apparent motion of the object in the sky, as
well as on the time scale of a few days, due to `weather'. This is
clearly a potential difficulty for planet detection, especially with
irregular time sampling, and can induce apparent variability on
sub-day and few-day time scales, as is observed here. If we
acknowledge that instrumental systematics, or any other cause, can
produce sudden RV jumps of order 20\,m\,s$^{-1}$ , we must treat with
extreme caution any claims to measure the semi-amplitudes of one or
more planetary companions at sub-m\,s$^{-1}$ precisions.

\subsection*{Possible sources of RV systematics}
\label{syst}

%Tally of level of HARPS systematics, and comparison with Keck and SOPHIE. Some reasonable generalisation. 

As discussed in the introduction, the precision of RV measurements degrades very rapidly and non-linearly at low SNR. This statement is based on experience accumulated over the past two decades \citep[see e.g.][]{Pon94,Pon97,Bou05,Pon05}, over a wide range of targets (from the Sun-like targets of transit surveys such as OGLE, SuperWASP, HATnet and CoRoT to Galactic Cepheids and extra-galactic red giants)  and instruments (CORAVEL, ELODIE and SOPHIE on the Observatoire de Haute Provence 1.93-m telescope, EMMI on the New Technology Telescope and HARPS on the La Silla Observatory 3.6-m telescope, and UVES on the Very Large Telescope). Other groups report similar experiences with the same or other instruments (VLT/UVES follow-up of SWEEPS candidates, \citealt{Sah06}, Keck/HIRES follow-up of transiting planets, M. Endl, J. Winn, priv. comm.). In all cases, the RV precision drops with decreasing SNR significantly faster than expected from photon noise and errors in the wavelength calibration alone, or from increased errors in fibre positioning for faint stars. For example, RV residuals for SOPHIE RVs typically depart from photon-noise expectations below ${\rm SNR} \sim 50$, and uncertainties at the level of 100\,m\,s$^{-1}$  or more appear when ${\rm SNR} \leq 20$. %The RV uncertainties become very large (up to 7\,km/s) long before the position of the CCF peak itself becomes uncertain, although formal errors remain in the range of a few hundred m/s down to $SNR=1$ (in a typical target/spectrograph combination with the Thorium/Argon wavelength calibration technique). 

%Explain that we are not talking about the wavelength calibration, but detector/background effects. 

If one wishes to speculate about the possible causes of this phenomenon, the most promising direction might be detector and background effects. The SNR that is typically quoted is the SNR per pixel at the centre of the central order in the echelle spectra used to derive the RVs. However, the SNR varies strongly from order to order, because the stellar continuum, the CCD response and the spectrograph throughput are all strongly wavelength dependent. This typically results in SNRs which are an order of magnitude lower in the blue half of the spectrum than in the red, particularly for the relatively cool targets of typical planet searches. As the overall SNR decreases, the signal in the blue half of the spectrum approaches the level of the various detector noise sources more rapidly than that in the red half of the spectrum. This has a disproportionate effect on the RV uncertainty because most of the thin metallic lines used to compute CCFs for solar-type stars are in the blue part of the spectrum. A 1\,m\,s$^{-1}$  RV accuracy corresponds to a displacement of the order of a thousandth of a CCD pixel. The wavelength calibration for individual lines is far from this accuracy, which can only be reached by combining -- effectively, averaging over -- thousands of lines. At low SNR, the weighing of this average is modified, resulting in non-random radial velocity offsets. Another source of non-random radial velocity uncertainty is the presence of background light around the object, particularly when due to diffused moonlight \citep[see e.g.][]{pep08}. This is  a well-known source of offsets in radial-velocity measurements for faint objects, and can be difficult to identify and correct at low contamination levels.

\subsection*{The case of HARPS}

HARPS is optimized to obtain extremely precise radial velocities from
high-SNR spectra, and it is mainly used in that mode, to search for
low-mass planets around relatively bright stars (typically
$V<8$). Assessing the possible amplitude of instrumental uncertainties
at lower SNR is difficult, because such data is relatively
sparse. However, HARPS was used for the brightest transit candidates
of the OGLE transit survey, at $V \sim 15$ \citep{Bou05}, as well as
part of the CoRoT and SuperWASP follow-up efforts ($11<V<16$ and
$9<V<12$ respectively). The data are currently public only for
confirmed planets.

Among these, a few cases are of particular interest. The planetary
nature of the transiting companion to CoRoT-1 ($V=13.6$) was
established with SOPHIE \citep{Bar08}, but relatively low-SNR HARPS
spectra were later acquired to measure the spectroscopic transit
(SNR=9--12) and these show residuals significantly larger than the
formal uncertainties (unpublished CoRoT mission follow-up data), up to
40\,m\,s$^{-1}$ . CoRoT-5 and CoRoT-4 ($V$=14.0 and 13.7) were also observed
with HARPS \citep{Mou08,Rau09} with SNRs of 50--60 in the red half of
the spectrum and 40--15 in the red and blue halves of the spectra,
respectively. The residual r.m.s. around the best-fit planet models
are 10\,m\,s$^{-1}$  and 18 m\,s$^{-1}$  respectively. WASP-6 is the only SuperWASP case
with a sufficient number of published measurements to study the
issue\footnote{Only 6 HARPS measurements are available for WASP-22, at
$V=12.0$. They are almost perfectly fitted by the best-fit planet
model (r.m.s. 4\,m\,s$^{-1}$ ), but little can be read into this, given the
that there are almost as few data points as degrees of freedom of the
fit.} \citep{Gil09}. Its $V$-magnitude is V=11.9, similar to that of
CoRoT-7. The HARPS measurements for WASP-6 have a residual r.m.s.\ of
12\,m\,s$^{-1}$  around the best-fit planet solution, while the mean formal
uncertainty is 7\,m\,s$^{-1}$ , indicating an unknown source of noise at the
10\,m\,s$^{-1}$  level. For some objects, the residuals could be due to an
intrinsic cause, such as stellar activity. The lightcurve of CoRoT-4
shows that it is fairly active.  However, there is no sign of
significant activity in either the spectroscopic indicators or
photometric monitoring of WASP-6, CoRoT-1 and CoRoT-5.

On the whole, there is a mounting body of evidence that unexplained variations at the 5--10\,m\,s$^{-1}$  level may exist in HARPS RVs for targets in the brightness range of CoRoT-7. Until their origin can be explained, the only safe course of action is to work on the assumption that the CoRoT-7 data itself may contain systematics at a similar level. 

We now examine the CoRoT-7 HARPS data for possible systematics in more
detail. One line of evidence in that direction comes from the bisector
span data. As shown in Figs.~\ref{fig1} and \ref{fig2},  these data
are entirely compatible with our simple spot model in its overall
evolution. However, contrarily to proxy photometry derived from the
CCF FWHM, the bisector span data display a point-to-point random
variation that is clearly larger than their uncertainties. The
bisector span is measured by comparing the radial velocity at the top
and the bottom of the CCF: its uncertainties are thus related to the
RV uncertainties in a rather straightforward way. On bright targets,
the bisector span uncertainty is expected -- and shown -- to be about
twice the RV uncertainty. In the case of CoRoT-7, the scatter of the
bisector span about the best-fit activity models is significantly
larger than twice the formal RV uncertainty, suggesting that the
latter are underestimated.

Finally, we look at the RV data themselves. The calculation of
the formal uncertainties assumes that the RV errors scale linearly
with the inverse of the spectrum SNR \citep{bou01}. On the other hand,
we suspect that the dependency of the errors on the SNR is stronger
than linear. To test this, we divided the RV residuals shown in
Fig.~\ref{fig3} by the SNR of the corresponding spectra. The scatter
of the lower-SNR data points is significantly larger than that of the
high-SNR points, by a much larger amount  than indicated by the
formal error bars. The same holds for the residuals relative to the
mean. This observation clearly indicates that the dependency of
the uncertainties on SNR is underestimated. It can be modelled by
assuming the presence of additional instrumental uncertainties with a
steep dependence on SNR, at the $\sim 5$\,m\,s$^{-1}$  level near the median
SNR, increasing to about $\sim 10$\,m\,s$^{-1}$  at the low end of the SNR
range.

\section{Evidence for and mass constraints on the planet(s)}

%\begin{figure}
%\includegraphics[width=\linewidth]{lomb.png}
%%\resizebox{12cm}{!}{\includegraphics{lomb.png}}
%\caption{Lomb-Scargle periodogram of the radial velocity residuals. Increasingly light colours indicate periodograms computed from subsets of the data with an increasingly high SNR threshold. The lightest includes only the third of the data with the best SNR.}
%\label{fig5}
%\end{figure}

\begin{figure}
\includegraphics[width=\linewidth]{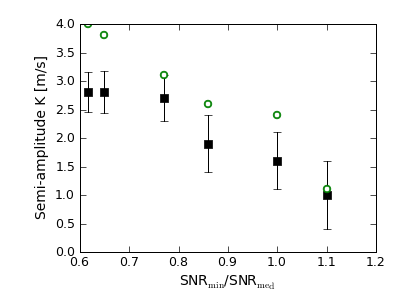}
%\resizebox{12cm}{!}{\includegraphics{plotK.png}}
\caption{Best-fit radial-velocity semi-amplitude $K_b$ for CoRoT-7b, as a function of the SNR threshold (expressed as a multiple of the median SNR). Closed symbols and error bars show the results with the RV contribution of stellar activity removed using the models, open symbols the values for the raw RV data. The monotonic trend in $K_b$ versus SNR threshold  suggests that SNR-dependent effects play a large role in the detected amplitude.}
\label{fig6}
\end{figure}

\subsection*{The CoRoT-7b RV signal and its amplitude}

The standard way of searching for the signature of planetary orbits in
RV data is to use a Lomb-Scargle or generalized periodogram
\citep{Hor86,Pre86,Zec09}. The periodogram of the HARPS RV data for
CoRoT-7 is highly complex. In the course of their pre-whitening
analysis, Q09 identify no less than eleven peaks, all of them highly
`significant' in the sense that they correspond to low formal false
alarm probabilities. However, on should bear in mind that the
false alarm probability expresses the probability that a given peak is
due to Gaussian white noise, but neither activity nor instrumental
noise are expected to be white or Gaussian. The very irregular
sampling of the data also implies that one or more of the peaks may
arise from signal at a single, apparently unrelated
frequency. Nonetheless, the main peaks in the RV periodogram are
clearly related to the signal from activity, being near the rotation
period, its harmonics, and their one-day aliases. There is also a peak
corresponding to the period of the transit signal detected in the
CoRoT photometry, $P_b=0.854$\,d.

%Fig.~\ref{fig5} shows the Lomb-Scargle periodogram of the raw RV data(no activity model removed) around the relevant frequency. The different lines illustrate the evolution of the periodogram as measurements derived from low-SNR spectra are progressively discarded,starting with the lowest SNR. 
Fig.~\ref{fig6} shows the semi-amplitude
of the best-fit sinusoid at the period and phase of the photometric
transit, as a function of SNR threshold, as measurements derived from
low-SNR spectra are progressively discarded, starting with the lowest
SNR. For the most stringent threshold (${\rm SNR}>1.15\,{\rm SNR}_{\rm
med}$), about one third of the measurements remain. We performed this
calculation with both the raw RV, and the residuals from our activity
models. The stellar rotation and planetary orbital frequencies are
widely separated, but the latter is close to the one-day alias of the
third harmonic of the former. As a result, it is not clear a priori
whether correcting for the activity signal improves the semi-amplitude
measurements, or on the contrary adds noise to them.

\begin{figure}
\includegraphics[width=\linewidth]{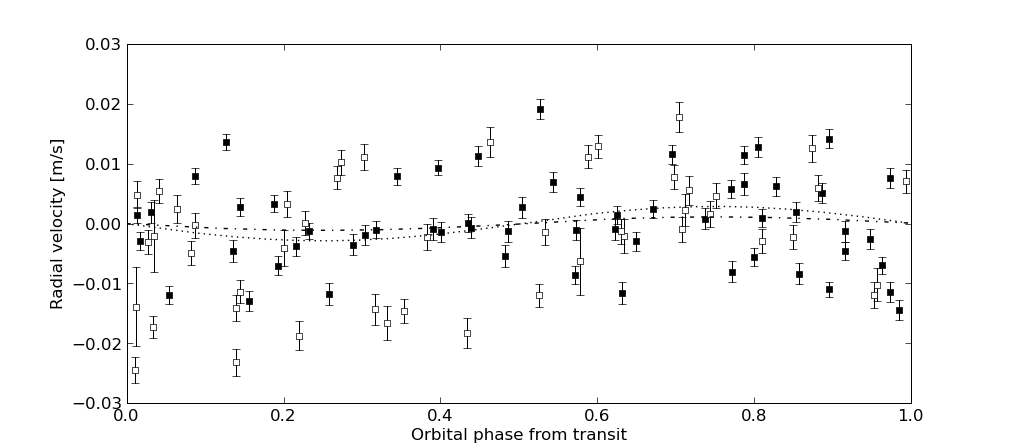}
\caption{ Radial-velocity data for CoRoT-7 phased to the transit signal, with
a typical realisation of our activity signal reconstruction subtracted. Open symbols indicate measurements
with SNR below the median. The lines show the best-fit orbit using all
measurements (dotted) and the higher-SNR measurements only
(dash-dotted).}
\label{phasedvr}
\end{figure}

Fortunately, the two methods give very similar results. In both cases,
the measured orbital semi-amplitude depends strongly on the SNR
threshold: including lower-SNR measurements favours a higher
value. Low-SNR measurements are more likely to be outliers (as their
formal uncertainties are underestimated), and would favour a higher
amplitude for all fitted features: the higher number of measurements
is offset by their poorer quality.  It is therefore not clear whether
the most reliable value of $K_b$ is the one derived from all the RV
measurements, or from only the best third or half. The monotonic trend
in $K_b$ versus SNR threshold suggests that SNR-dependent effects play
a large role in the detected amplitude. Figure~\ref{phasedvr}
shows the radial-velocity data corrected by one of our variability
models and phased to the transit signal, together with the best-fit
Keplerian orbit with and without an SNR cut.

To evaluate the effect of the steeper rise in the total uncertainty at
the lower end of the SNR range, we re-calculate the constraints on $K$
in the following way: we add a new term to the radial-velocity
uncertainties, with a quadratic rather than linear dependence on the
inverse of the signal-to-noise ratio. We set the magnitude of this
term so that the reduced $\chi^2$ of the residuals of the best-fit
model (including activity and the orbit of CoRoT-7b) is unity. The
resulting value of $K$, averaged over all spot models, is
$K=1.6$\,m\,s$^{-1}$. We then estimate the uncertainty on this value
using a bootstrap on the individual measurements. This provides an
error estimate that is independent on the assumptions regarding the
nature of the sources of uncertainties and their distribution. We find
$\sigma_K=1.3$\,m\,s$^{-1}$. The corresponding mass range for CoRoT-7b
is $m_p = 2.3 \pm 1.8$ M$_{\rm Earth}$.

\begin{figure}
\resizebox{8cm}{!}{\includegraphics{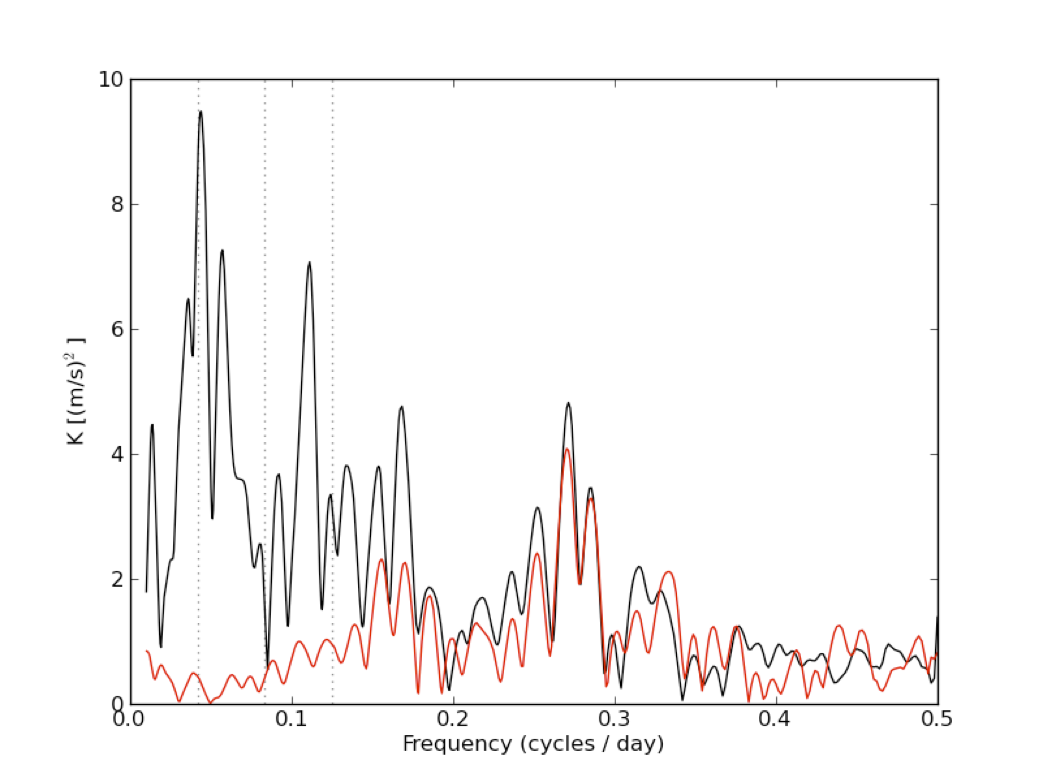}}
\caption{Sine-fitting periodogram of the HARPS RV data before (black) and after (red) applying the adaptive harmonic filter of Q09. The y-axis shows the amplitude of the best-fit sinusoid at each trial frequency (the phase and zero-point are free). The frequency range covers the full Nyquist sampling interval. The amplitude at 0.17 cycles/day (the one-day alias of the period of CoRoT-7b) is reduced from $\sim 5$ to $\sim 2$\,m\,s$^{-1}$. }
\label{figsa}
\end{figure}

Our uncertainty is significantly larger than that quoted by Q09
($0.8\,$m\,s$^{-1}$). Q09 estimate that their pre-whitening filter
induces a systematic uncertainty of up to 1\,m\,s$^{-1}$, but they do
not incorporate this in their final uncertainty. Similarly, they noted
that the adaptive harmonic filter significantly reduces the amplitude
of the signal at the period of CoRoT-7b, and correct for this by
multiplying both the semi-amplitude measured after applying that
filter and its formal uncertainty by a factor of two. For comparison,
we have reproduced the adaptive harmonic filtering used in Q09. The
result is shown on Fig.~\ref{figsa}. We find that it reduces the signal at the period of
CoRoT-7b by 2\,m\,s$^{-1}$ (from $\sim 4$ to $\sim
2$\,m\,s$^{-1}$). Since there is no compelling way of assessing how
much of the signal that was filtered out arose from activity, and how
much was of planetary origin, we estimated that the systematic
uncertainty induced by the Q09 filtering process is $\sim
2$\,m\,s$^{-1}$.

\subsection*{Additional planets?}

The RV data are dominated by strong and highly complex
activity-induced jitter. In addition, we have shown that it also
contains unrecognised features, probably instrumental in origin, with
unknown spectral characteristics, and amplitudes in the range
5--10\,m\,s$^{-1}$. Given this, we believe it is not warranted to
assess the possible presence of the low-mass planets at $P_b=3.69$\,d
and $P_c=9.03$\,d considered by Q09 and H10.

We did investigate the possible detection of a signal from CoRoT-7b,
because the transits provide a-priori knowledge of its period and
phase, placing relatively strong constraints on its RV
signature. However, no such constraints exist for non-transiting
planets. The combined effect of activity, additional uncertainties in
the low-SNR data, and irregular sampling, could easily give rise to
periodogram features of amplitude similar to the semi-amplitudes
expected for Neptune-mass planets around CoRoT-7 at periods  in
the $3$--$10$\,d range. We recall that typical timescales for
weather-induced effects fall in that range, and that both we and
\citet{Lan10b} find evidence for differential rotation as well as
rapid active region evolution, which further the range of
periodicities at which activity can be expected to contribute to
the signal.

\begin{figure}
\includegraphics[width=\linewidth]{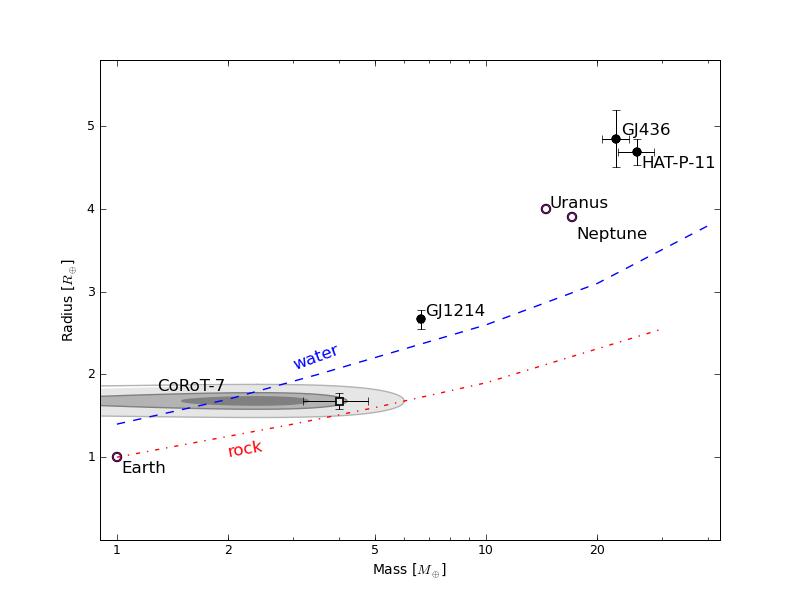}
%\resizebox{12cm}{!}{\includegraphics{MRplot.png}}
\caption{  Mass-radius diagram for medium-size planets, showing known transiting exoplanets and Solar System planets. The lines illustrate the approximate locus of structure models for three extreme in composition, pure rock, pure water ice and pure H/He gas. The square with error bars indicate the location of CoRoT-7b in Q09. The revised location of CoRoT-7b is shown, with the 50\%, 75\% and 95\% probability contours indicated by gray shades.}
\label{fig7}
\end{figure}

\subsection*{Is CoRoT-7b detected in RV?}

The detection of a planetary RV signal at the period of CoRoT-7b is
only significant at the $1.2\,\sigma$ level. Nevertheless, the RV
periodogram does contain a peak at the photometric period, and the RV
orbital solution is in phase with the transit solution of the CoRoT
photometry. While it is difficult to quantify this evidence, it does seem to
add credibility to the hypothesis of the planet existence and the
likelihood of a significant RV planet detection.
On the other hand, it is important to keep in mind that the observations of CoRoT-7 were not collected in a `blind' way. Instead, they were obtained with the specific aim of searching for a low-mass planet signal at the period and phase of the transits. This induces a subjective bias in the observations: additional HARPS measurements were scheduled until a signal of amplitude deemed sufficient was apparent in the noise. This `noise-enhanced detection' effect makes it difficult to apply statistical arguments to assess the phase coincidence.

To summarise, the RV data is at best a marginal confirmation of the planetary nature of the eclipsing companion. We note that the rest of the follow-up effort reported by \citet{Leg09} excluded most, but not all, blended eclipsing binary scenarios. A blend would also give rise to sinusoidal variations at the period and phase of the transits \citep{Kon04}. There are several cases in the literature where RV confirmation of transiting planets has proved tricky and announcements were made on the basis of marginal RV detections. Some of these were later shown to be spurious \citep{Drei03,Kon03}, while some are widely considered to remain unsolved (SWEEPS-4 and 11, \citealt{Sah06}). In this context, it is difficult to be confident that CoRoT-7b meets the requirements for planetary confirmation.

%Were it not such a newsworthy object, CoRoT-7b would not be considered as meeting the requirements for planetary confirmation in the context of transit survey. 

\subsection*{The nature of CoRoT-7b}

Let us for now take the measurement of the RV semi-amplitude of CoRoT-7b at face value, and estimate the resulting constraints on its mass, $m_b$.

Figure~\ref{fig7} shows the position of the equiprobability contours in the mass-radius diagram corresponding to our results. The most likely value for the density of the planet corresponds to a water/ice composition \citep[or a rocky core and a large H/He envelope, as the two are degenerate, as pointed out by e.g.][]{Rog10}. This composition is marginally favoured (at the $\sim 1\,\sigma$ level) over a rocky planet, or a much lighter H$_2$O or Hydrogen-rich planet. As the RV detection is marginal, negligible masses (undetectable with the present data) cannot be firmly excluded. 

Compared to other extra-solar planets in the 1--$20\,M_{\rm Earth}$ range, the updated position of CoRoT-7b raises an interesting possibility, which was seemingly excluded by the Q09 value. All the extra-solar planets in this mass range, including CoRoT-7b, seem most consistent with a composition dominated by water ice, similar to Neptune and Uranus. The same positions are consistent with a primarily rocky composition with a large gas envelope, but fine-tuning would be required for such planets to have mean densities identical to those expected from a Uranus/Neptune-like composition. This suggests that those transiting planets, now on very close-in orbits, may all have been formed beyond the snow line. CoRoT-7b could be a 'mini-Neptune' rather than a 'Super-Earth', although such speculation remains subject to the proviso that the confirmation of its planetary nature is not as firm as that of the other planets in the low-mass transiting sample.

%
%4- Mass constraints on planet if any

%Raw amplitude. Amplitude without fishy data. Uncertainty estimates: Suz, boot, jumps --> ~60%

%Revise position, log-mass prior. 

%Discuss possibility of non planet: suggestive that there is power at the right period and epoch. ~2-sigma result. Good, but would not be considered confirmation (examples. OGLE-TR-3, SWEEPS-4 or 11). Models have to keep this in mind.

%Rocky nature (K>3.5) not supported by data to any satisfactory level. Ice density more likely according to our analysis, or rock/gas, see Seager.

%Give value and error bars. Place in plot. Discuss scientific implications: up to now, all light planet radius measurements are compatible with an out-of-snowline formation. That is striking. CoRoT-7 no longer a credible exception.

%\begin{figure}
%%\resizebox{12cm}{!}{\includegraphics{plotmass.png}}
%\includegraphics[width=\linewidth]{plotmass.png}
%\caption{Likelihood as a function of mass for CoRoT-7b (with $K=1.7 \pm1.3$\,m/s).}
%\label{fig8}
%\end{figure}

\section{Conclusions}

We have performed a detailed analysis of the HARPS observations of CoRoT-7 published by Q09, constructing a realistic model of the activity-induced stellar signal, making use of all the available data, and exploring the possibility of errors beyond the formal RV uncertainties. 

We find that the signal from stellar activity during the HARPS observations can be robustly modelled by dark spots rotating on the surface of the star, using the CCF width and bisector information to constrain the model. We also find clear evidence of systematics in the RV data, in the form of large jumps (of order 20\,m\,s$^{-1}$) which are significantly larger than the formal uncertainties and are explained neither by activity nor by the putative planetary signal. As these unaccounted-for effects depend strongly on the SNR of the spectra, we attribute them to instrumental uncertainties operating in the mid-SNR regime, and which are also seen in other stars observed at similar SNR with HARPS and other high-precision RV spectrographs.

Allowing for the SNR-dependent uncertainties, we estimate the semi-amplitude of the signal at the period and phase of CoRoT-7b to be $1.6\pm1.3$\,m\,s$^{-1}$, a detection at the $1.2\,\sigma$ level. This value corresponds to a companion mass of $m_b=2.3 \pm 1.8\,M_{\rm Earth}$, with the 95\% confidence interval encompassing the full 0--5$\,M_{\rm Earth}$ range. Given the presence of strong variations of stellar and instrumental origin with unknown spectral characteristics, we argue that the data cannot be used to search for additional (non-transiting) planets in the $3$--$10$\,d period range, and that claims of the detection of such planets ('CoRoT-7c' and 'CoRoT-7d') does not stand scrutiny. 

We conclude that the data provides at best marginal evidence for the
presence of a planet in orbit around CoRoT-7 at the period of the
transits detected in the CoRoT data. If the planetary hypothesis is
adopted, the data allow for a range of compositions, and favour a
somewhat lower mean density than previously stated, implying that the
rocky nature of CoRoT-7b is far from certain.

In the future, the Kepler and CoRoT missions are expected to yield
more terrestrial planet candidates. Our analysis of the case of
CoRoT-7b demonstrates the importance of securing simultaneous RV and
photometry follow-up especially for those cases of active host
stars. The basic many-spot model which we used can be further
elaborated in several directions, such as introducing Bayesian
analysis techniques or more sophisticated modelling of the
stellar surface. We suggest it may prove to be a basic item in the
analysis toolbox for those cases, both in cases where simultaneous
photometry exists or when proxy photometry through the CCF has to be
used (Aigrain et al., in prep.)

More RV measurements are needed to confirm the planetary nature of CoRoT-7b beyond reasonable doubt and to improve observational constraints on its mass to a level where it can be usefully compared to theoretical models. This could be done on a larger telescope, for instance Keck/HIRES, in a reasonable time and at higher signal to noise -- thus testing the SNR-dependence of the RV uncertainties, and hopefully circumventing the issue. If systematics at the level of a few m\,s$^{-1}$ can be excluded, and a reliable brightness indicator collected at the time of the observations, a few measurements per night during a few nights should be sufficient to measure the short-timescale component of the signal with reasonable accuracy. 

In the meantime, we caution that models building on the rocky nature of CoRoT-7b may be built on sand.

\section*{Acknowledgements}

We wish to acknowledge the support of a STFC Advanced Fellowship (FP), an STFC Standard Grant ST/G002266/1 (SA),and  the Israel Science Foundation / Adler Foundation for Space Research Grant No. 119/07 (SZ).

\bibliographystyle{mn2e}

\label{lastpage}

\end{document}